\documentclass[aps,twocolumn,nofootinbib,superscriptaddress]{revtex4}

\usepackage{graphicx}  
\usepackage{dcolumn}   
\usepackage{bm}        
\usepackage{amssymb,amsfonts,amsmath,physics}   
\usepackage{cancel}
\usepackage{mathtools}
\usepackage{slashed}
\usepackage[dvipsnames]{xcolor}

\usepackage{tikz-feynman}
\tikzfeynmanset{compat=1.1.0}
\usetikzlibrary{arrows}
\tikzset{	
	vertex/.style={circle,draw, minimum size=1.5em},	
	edge/.style={->,> = latex'}	
}
\usepackage[bookmarks, breaklinks, colorlinks,urlcolor=blue, citecolor=red, 
linkcolor=blue]{hyperref}
\usepackage[normalem]{ulem}

\hypersetup{
	colorlinks=true,
	linkcolor=Red,
	filecolor=magenta,
	urlcolor=Blue,
	citecolor=PineGreen}

\newcommand{\be}{\begin{eqnarray*}}
	\newcommand{\ee}{\end{eqnarray*}}

\newcommand{\bee}{\begin{eqnarray}}
	\newcommand{\eee}{\end{eqnarray}}
\newcommand{\beeq}{\begin{equation}}
	\newcommand{\eeq}{\end{equation}}

\usepackage{xspace}

\newcommand{\ba}{\begin{array}}
	\newcommand{\ea}{\end{array}}
\newcommand{\bd}{\begin{displaymath}}
	\newcommand{\ed}{\end{displaymath}}
\newcommand{\besub}{\begin{subequations}}
	\newcommand{\eesub}{\end{subequations}}
\newcommand{\bea}{\begin{eqnarray}}
	\newcommand{\eea}{\end{eqnarray}}

\newcommand{\sla}[1]{/\!\!\!#1}

\def\q2 {q^2}

\usepackage{tikz}
\usetikzlibrary{arrows,shapes}
\usetikzlibrary{trees}
\usetikzlibrary{matrix,arrows} 				
\usetikzlibrary{positioning}				
\usetikzlibrary{calc,through}				
\usetikzlibrary{decorations.pathreplacing}  
\usepackage{pgffor}							

\usetikzlibrary{decorations.pathmorphing}	
\usetikzlibrary{decorations.markings}
\tikzset{
	vector/.style={decorate, decoration={snake}, draw},
	provector/.style={decorate, decoration={snake,amplitude=2.5pt}, draw},
	antivector/.style={decorate, decoration={snake,amplitude=-2.5pt}, draw},
	fermion/.style={draw=black, postaction={decorate},
		decoration={markings,mark=at position .55 with {\arrow[draw=black]{>}}}},
	fermionbar/.style={draw=black, postaction={decorate},
		decoration={markings,mark=at position .55 with {\arrow[draw=black]{<}}}},
	fermionnoarrow/.style={draw=black},
	gluon/.style={decorate, draw=black,
		decoration={coil,amplitude=4pt, segment length=5pt}},
	scalar/.style={dashed,draw=black, postaction={decorate},
		decoration={markings,mark=at position .55 with {\arrow[draw=black]{>}}}},
	scalarbar/.style={dashed,draw=black, postaction={decorate},
		decoration={markings,mark=at position .55 with {\arrow[draw=black]{<}}}},
	scalarnoarrow/.style={dashed,draw=black},
	electron/.style={draw=black, postaction={decorate},
		decoration={markings,mark=at position .55 with {\arrow[draw=black]{>}}}},
	bigvector/.style={decorate, decoration={snake,amplitude=4pt}, draw},
}

\tikzstyle{block} = [draw, rectangle, 
minimum height=3em, minimum width=6em]

\begin{document}
	
	\title{Leptogenesis with Majoron Dark Matter}
	
	\author{Stephen F. King}
	\email{S.F.king@soton.ac.uk}
	\affiliation{School of Physics and Astronomy, University of Southampton,
		Southampton SO17 1BJ, United Kingdom}
	
	\author{Soumen Kumar Manna}
	\email{skmanna2021@gmail.com}
	\affiliation{Department of Physics, Indian Institute of Technology Guwahati, Assam-781039, India}
	
	\author{Rishav Roshan}
	\email{r.roshan@soton.ac.uk}
	\affiliation{School of Physics and Astronomy, University of Southampton,
		Southampton SO17 1BJ, United Kingdom}
	
	\author{Arunansu Sil}
	\email{asil@iitg.ac.in}
	\affiliation{Department of Physics, Indian Institute of Technology Guwahati, Assam-781039, India}
	
	\begin{abstract} 
		We discuss a model of neutrino mass based on the type I seesaw mechanism embedded in a spontaneously broken global lepton number framework 
		with a $Z_2$ symmetry. 
		We show that the resulting Majoron is a viable freeze-in dark matter candidate. Two right-handed neutrinos are assumed to have dominant off-diagonal masses suggesting resonant leptogenesis as the origin of baryon asymmetry of the Universe.  Explicit higher dimensional lepton number violating operators, are shown to play a crucial role in simultaneously controlling both the Majoron production in the early Universe and the right handed neutrino mass splitting relevant for resonant leptogenesis. We perform a combined analysis of Majoron dark matter and leptogenesis, discussing the relative importance of self energy and vertex contributions to CP asymmetry, and explore the parameter space, leading to an intricate relation between neutrino mass, dark matter and baryon asymmetry. 
		
	\end{abstract}
	\maketitle


	\section{Introduction}
	\label{sec.introduction}
	Although the origin of neutrino mass and mixing is unknown, the type I seesaw mechanism \cite{Minkowski:1977sc,Mohapatra:1979ia,Schechter:1980gr} remains an attractive possibility by virtue of its most economic construction beyond the Standard Model (SM). In its minimal version, only two massive right-handed neutrinos (RHN) are sufficient to account for the neutrino oscillation data \cite{King:1999mb}. Additionally, if these sterile RHNs are associated with an (approximate) off-diagonal mass matrix leading to two (approximately) degenerate Majorana mass eigenstates \cite{King:2002nf}, they can naturally be part of a resonant leptogenesis \cite{Pilaftsis:2003gt,Ibarra:2003up} framework in order to explain the matter antimatter asymmetry of the Universe. 
	
	However, the Majorana masses of these RHNs clearly violate the lepton number ($L$). Therefore, the origin of masses of the RHNs can be attributed to the spontaneous breaking of a global lepton number symmetry $U(1)_{L}$ by a SM singlet complex scalar field ($\Phi$), coupled to RHNs, such that its vacuum expectation value ($vev$) generates the Majorana mass matrix of RHNs. Such a spontaneous breaking of the global $U(1)_{L}$ then results in a massless Nambu-Goldstone boson called Majoron ($\chi$) \cite{Chikashige:1980qk, Chikashige:1980ui, Gelmini:1980re}. Allowing small explicit (soft) $U(1)_{L}$ breaking operators, the Majoron acquires a small mass and provides a possible dark matter (DM) candidate \cite{Akhmedov:1992hi,Rothstein:1992rh,Berezinsky:1993fm,Bazzocchi:2008fh,Gu:2010ys, Frigerio:2011in,Shakya:2018qzg,Lattanzi:2013uza, Queiroz:2014yna} as its stability can be ensured by the $vev$ suppressed interactions with other fields. 
	
	The  freeze-out mechanism for such Majoron DM production has been studied via a Higgs portal coupling
	\cite{Frigerio:2011in,Queiroz:2014yna} (see also \cite{Cline:2013gha}) with strong constraints from XENON-1T 
	\cite{Queiroz:2014yna,Cline:2013gha} suggestive of a heavy Majoron beyond TeV, which however
	fails to meet the stability criteria, as it would decay into two light neutrinos via active-sterile neutrino mixing \cite{Minkowski:1977sc,Mohapatra:1979ia,Schechter:1980gr}. On the other hand, the freeze-in mechanism \cite{Hall:2009bx} for Majoron DM production has also attracted a lot of attention in recent days since its coupling 
	to SM particles is suppressed by the $U(1)_{L}$ breaking scale \cite{Escudero:2021rfi}, and hence it presents itself as a natural candidate for a feebly interacting massive particle (FIMP) DM. In particular, with the help of lepton number breaking (soft) Higgs portal coupling, Majorons can be produced from the decay of the SM Higgs \cite{Frigerio:2011in,Garcia-Cely:2017oco,Brune:2018sab}. However this mechanism is tightly constrained, leading to a fine-tuned Majoron mass $m_\chi \sim$ 3 MeV \cite{Frigerio:2011in,Garcia-Cely:2017oco,Brune:2018sab}\footnote{Alternatively non-thermal production of TeV-scale Majoron via UV freeze-in has been considered \cite{Abe:2020dut}}. To relax this, a new production mechanism for Majoron as a FIMP-type DM was proposed by some of us \cite{Manna:2022gwn} based on a dimension-5  $U(1)_{L}$ breaking operator instead of Higgs portal couplings, resulting in an extended range of Majoron as DM having mass ranging from $\mathcal{O}(\rm{keV})$ to $\mathcal{O}(\rm{GeV})$ \cite{Manna:2022gwn} accessible by neutrino experiments Borexino\cite{Borexino:2010zht}, KamLAND\cite{KamLAND:2011bnd}, and Super-Kamiokande (SK)\cite{Super-Kamiokande:2002exp}, and $\gamma$-ray observations INTEGRAL \cite{Boyarsky:2007ge}, COMPTEL/EGRET \cite{Yuksel:2007dr}, and Fermi-LAT \cite{Fermi-LAT:2015kyq}.
	
	In this paper, we consider the above scenario of Majoron as FIMP-type DM with resonant leptogenesis. 
	To be explicit, we discuss a concrete $U(1)_{L}$ symmetric model of neutrino mass based on the type I seesaw mechanism, involving two right-handed neutrinos and a $U(1)_L$ breaking scalar field having $Z_2$ odd parity, coupled to the RHNs at tree level. The construction leads to off-diagonal nonzero entries in the RHN mass matrix that result into two exactly degenerate RHNs. 
	Then, the inclusion of dimension-5 $U(1)_{L}$ breaking terms not only breaks such a degeneracy but also opens up new annihilation channels of RHNs to Majorons, thereby making Majoron production possible via freeze-in scenario \footnote{{A similar concept was explored in \cite{Bhattacharya:2021jli}, where the soft breaking of lepton number symmetry played a key role in the freeze-in production of dark matter.}} in line with \cite{Manna:2022gwn}. Note that the same splitting that removes the degeneracy of RHNs turns out to be also instrumental in producing the matter-antimatter asymmetry via resonant leptogenesis.

	We perform a combined analysis of Majorons as FIMP type dark matter and resonant leptogenesis and establish the parameter space which implies a relation involving neutrino mass, dark matter and the baryon asymmetry of the Universe. In doing so, we realize that contrary to the usual expectation of resonant leptogenesis happening near the TeV scale or so, here we end up a relatively high scale resonant leptogenesis scenario in order to be consistent with the Majoron DM parameter space, which can also be of interest for further investigation. 
	
	The layout of the remainder of the paper is as follows. In Section~\ref{sec.setup}, we define {the minimal type I seesaw setup that involves two RHNs and the lepton number symmetry breaking scalar field $\Phi$, featuring both global $U(1)_L$ conserving and breaking terms. 
		In Section~\ref{DM}, we discuss Majoron FIMP DM, focusing on the explicit $U(1)_L$ breaking non-renormalisable interactions that drive RHNs annihilation into Majorons and analyzing the Boltzmann equations leading to successful DM production.} 
	In Section~\ref{leptogenesis}, we discuss resonant leptogenesis in the considered model and calculate the lepton asymmetry for a benchmark point consistent with those required for Majoron FIMP DM.
	{In Section~\ref{parameterspace}, we discuss in detail the parameter space compatible with Majoron FIMP DM and resonant leptogenesis, highlighting the role of non-renormalizable $U(1)_L$-breaking operator common to both effects and the CP asymmetry contributions from self energy and vertex correction terms.} Section~\ref{conclusions} concludes the paper.
	

	\section{The type I seesaw Set-up}
	\label{sec.setup}
	
	In this section, we establish the type I seesaw framework to accommodate Majoron as a dark matter and to achieve resonant leptogenesis. Similar to the original Majoron model, we extend the SM by including two singlet right handed neutrinos 
	$(\mathcal{N}_{1,2})$ and a SM singlet complex scalar field $(\Phi)$, charged under a global $U(1)_L$ and an additional 
	$Z_2$ symmetry. While all the SM fields and one of the RHNs $\mathcal{N}_{1}$ are even under this $Z_2$, $\mathcal{N}_{2}$ and $\Phi$ carry odd $Z_2$ parity, {as shown in table \ref{tab-charge}.} 
	\begin{table}[hbt]
		\begin{center}
			\vskip 0.5 cm
			\begin{tabular}{|c|c|c|c|}
				\hline
				Symmetries	&	$\Phi$ & $\mathcal{N}_1$ &  $\mathcal{N}_2$   \\
				\hline
				$U(1)_L$       &  $-2$     &  1 & 1  \\                 
				\hline
				$Z_2$    & $-$ &+ & $-$  \\                 
				\hline
			\end{tabular}
		\end{center}  
		\caption{Beyond the SM particles and their charges under global $U(1)_L$ and $Z_2$ symmetries.}
		\label{tab-charge}
	\end{table}
	As a result, though the field contents are the same, our setup is relatively different than the original Majoron model in terms of the renormalizable part of the Lagrangian. Here, we consider the following part of the type I seesaw Lagrangian, which respects both 
	$U(1)_L$ and $Z_2$ symmetry, 
	\beeq
	-\mathcal{L}_{\rm SC}\supset  \frac{f}{2}\Phi \overline{\mathcal{N}_1^C}\mathcal{N}_2+y_{\alpha 1}\overline{L_\alpha} \tilde{H}\mathcal{N}_1 +h.c.
	\label{eq:seesaw}
	\eeq
	Here $H$ is the SM Higgs doublet and $y_{\alpha 1}$ defines the coupling corresponding to the neutrino Yukawa interaction involving $\alpha=e,\mu,\tau$ lepton doublets with RHN $\mathcal{N}_1$. The other RHN does not contribute to the neutrino 
	Yukawa interaction at this renormalizable level due to $Z_2$ charge assignment of the fields involved. 
	Once the $U(1)_L$ symmetry is spontaneously broken $(\Phi=\left({\phi+v_\phi+i\chi}\right)/{\sqrt{2}})$, $\Phi$ field acquires a vacuum expectation value which leads to the RHN mass matrix (in flavor basis of RHNs) as 
	\bea
	M_R = \begin{pmatrix}
		0 & M \\
		M & 0
	\end{pmatrix},
	\eea
	with $M = f v_{\phi}/\sqrt{2}$. 
	It then follows that such a construction leads to an exactly degenerate pair of RHNs (barring a relative phase) having mass $M_1 = M_2 =  fv_\phi/\sqrt{2}$ at the renormalizable level. 
	
	Upon the spontaneous symmetry breaking (SSB) of global $U(1)_L$, the massless Nambu-Goldstone Boson (namely Majoron $\chi$) also results in presence of the scalar potential involving $\Phi$ and $H$ as follows
	\beeq
	V(H,\Phi)=V_H-\frac{\mu_\Phi^2}{2}|\Phi|^2+\frac{\lambda_\Phi}{2}|\Phi|^4
	+\lambda_{H\Phi}|H|^2|\Phi|^2,\eeq
	where $V_H=-\mu_H^2 H^\dagger H+\lambda_H (H^\dagger H)^2$ is the potential of SM Higgs and $\lambda_{H\Phi}$ implies the Higgs portal coupling. {Furthermore, we shall consider the portal coupling $\lambda_{H\Phi}$ to be small enough so as no 
		neglect the mixing between the CP even part of $H$ and $\Phi$}.  Followed by the SSB of $U(1)_L$ and subsequently the electroweak symmetry breaking (EWSB), the masses of the scalar fields take the form as
	\beeq
	m_h^2\simeq 2\lambda_H v^2,~~m_\phi^2\simeq \lambda_\Phi v_\phi^2,~~m_\chi^2=0,
	\eeq
	where $m_h$ is usual SM Higgs boson mass, 125 GeV \cite{CMS:2012qbp}, while $\chi$ remain massless. 
	
	\subsection{Explicit $U(1)_L$ breaking terms of higher order}
	
	The massless Majorons would get mass once the explicit lepton number symmetry breaking terms are introduced making them massive pseudo-Goldstone boson (pNGB) \cite{Lusignoli:1990yk}. For example, inclusion of soft explicit symmetry breaking term as 
	\beeq
	-\mathcal{L}_{\rm{soft}}=-\frac{m^2}{4}(\Phi^2+\Phi^{*2}),
	\label{eq:chi-mass}
	\eeq
	breaks $U(1)_L$ to a residual $Z_2$ symmetry (respected by $\Phi$) and induces a Majoron mass $m_\chi^2=m^2$. The term is a soft-breaking one, and natural, in a sense that, in the limit $m\to 0$, the symmetry of the framework is enhanced.

	The $U(1)_L$ symmetry, being a global one, is expected to be explicitly broken also by gravity effect at the Planck scale \cite{Giddings:1988cx,Coleman:1988tj,Rey:1989mg,Abbott:1989jw,Barbieri:1979hc,Akhmedov:1992hh} or even at a lower cut-off scale $\Lambda$ in the context of weak gravity conjecture as in \cite{Draper:2022pvk,Cordova:2022rer}. To be specific, here, we consider explicit $U(1)_L$ breaking terms at dimension-5 level involving $\Phi$ and two RHNs (while keeping the $Z_2$ symmetry intact) as
	\beeq
	\begin{multlined}
		-\mathcal{L}_{\sla{L}}= \frac{c_1}{2\Lambda}{\left[\Phi^2+(\Phi^*)^2\right]\overline{\mathcal{N}_1^C}\mathcal{N}_1}+\\\frac{c_2}{2\Lambda}{\left[\Phi^2+(\Phi^*)^2\right]\overline{\mathcal{N}_2^C}\mathcal{N}_2}
		+\frac{y_{\alpha 2}}{\Lambda}\overline{L_\alpha} \tilde{H}\mathcal{N}_2 (\Phi+\Phi^*)+h.c..
	\end{multlined}
	\label{eq:dim5-explicit}
	\eeq	
	{While the first two terms contribute to the diagonal elements of the RHN mass matrix as well as production of Majorons after $U(1)_L$ is spontaneously broken, the third term in the RHS of Eq. \ref{eq:dim5-explicit} signifies the Yukawa interaction of the second RHN $(\mathcal{N}_2)$. The uniform coupling of this interaction with $\Phi$ and $\Phi^*$ is ensured in presence of a CP symmetry \cite{Gross:2017dan} under which $\Phi \rightarrow \Phi^*$. The case of non-uniform coupling in considered in appendix \ref{apndx_A}. Note that other terms in Eqs. \ref{eq:chi-mass}, \ref{eq:dim5-explicit} are also consistent with such a symmetry consideration. As a result, the Majoron is not involved in this Yukawa interaction.} The involvement of this Yukawa coupling is crucial in making the light neutrino mass matrix to be consistent with neutrino oscillation data. 
	The other standard dimension-5 operator contributing to the light neutrino mass, $\frac{\delta_{ij}}{\Lambda}L_i H L_j H$, might also be present \cite{Weinberg:1979sa}. However, assuming $\delta_{ij}$ is sufficiently small, we exclude this contribution from the analysis without any loss of generality. Also, the presence of such $Z_2$-odd charge of $\Phi$, even for the $U(1)_L$ breaking contribution, ensures that terms such as $\Phi^3 |H|^2$ do not appear at the dimension 5 level. It can be noted that the origin of the soft breaking term in Eq. \ref{eq:chi-mass} can be considered as reminiscent of an explicit $U(1)_L$ breaking terms of higher order, e.g. $\beta \frac{|\Phi|^4}{\Lambda^2}(\Phi^2+\Phi^{*2})$. 
	
	\subsection{Effect of explicit $U(1)_L$ breaking on RHN mass}
	
	After the SSB of $U(1)_L$, the relevant Lagrangian inclusive of the explicit lepton number breaking contribution for RHN mass, in the flavor basis of RHN $(\mathcal{N}_1~\mathcal{N}_2)^T$, is given by 
	\beeq
	\mathcal{L}\supset \frac{1}{2}\begin{pmatrix} \overline{\mathcal{N}_1^C} & \overline{\mathcal{N}_2^C}  \end{pmatrix}\begin{pmatrix}
		{\kappa_1} & M \\
		M & {\kappa_2}
	\end{pmatrix} \begin{pmatrix} \mathcal{N}_1 \\  \mathcal{N}_2 \end{pmatrix},
	\eeq
	where {$\kappa_i =c_i {v_\phi^2}/{\Lambda}$} corresponds to the diagonal contribution to the mass matrix of RHNs arising {from the first two dimension-5 explicit breaking operators of Eq. \ref{eq:dim5-explicit}}. {Considering 
		$c_1 = c_2 = 1$ for simplification purpose (hence, $\kappa_{1,2} = \kappa$ say)}, we diagonalize this RHN mass matrix $(M_R)$ through a unitary operator $V$ following the relation $ V^* M_R V^\dagger = M_R^{d}$, where 
	\beeq
	M_R^d = {\rm{diag}}[M+\kappa, M-\kappa],
	\label{RHN mass}
	\eeq
	is the diagonal RHN mass matrix (barring a phase) in the basis of the RHNs: $(N_{R_1}~N_{R_2})^T$, connected to the flavor basis as: $\mathcal{N}_1=\frac{1}{\sqrt{2}}(N_{R_1}+i N_{R_2})~~{\rm and}~~\mathcal{N}_2=\frac{1}{\sqrt{2}}(N_{R_1}-iN_{R_2})$. Hence, the dominant contribution to RHN mass comes from the renormalizable part of the Lagrangian while the small splitting $(\kappa)$ emerges from the dimension-5 operators. These nearly degenerate RHNs would take part in resonant leptogenesis as we will discuss in a later section.

	
	

	\subsection{Neutrino mass}
	
	The Lagrangian for the neutrino sector is given by 
	\beeq
	\begin{multlined}
		-\mathcal{L}_{\nu} = y^{\nu}_{\alpha i}\overline{L_\alpha} \tilde{H}N_{R_i} +
		\frac{f v_\phi}{2\sqrt{2}} (\overline{N_{R_1}^C} N_{R_1} + 
		\overline{N_{R_2}^C} N_{R_2}) \\+
		\frac{\kappa}{2} (\overline{N_{R_1}^C} N_{R_1}-\overline{N_{R_2}^C} N_{R_2}) + h.c.,
		\label{eq:neutrino}
	\end{multlined}
	\eeq
	defined in the RHNs as well as charged lepton mass diagonal bases. Note that $y^\nu$ contributing to the Dirac mass of the neutrinos via $m_D=y^\nu v/\sqrt{2}$ after the EWSB, appears here as an effective Yukawa, originating from both the tree level and higher order Yukawas (due to the rotation of the RHN flavor to mass basis) as 
	\beeq
	\begin{multlined}
		y^\nu_{\alpha 1} = \frac{1}{\sqrt{2}}y_{\alpha1}+\frac{v_\phi}{\Lambda}y_{\alpha 2},\\
		y^\nu_{\alpha 2} = \frac{i}{\sqrt{2}}y_{\alpha1} - i\frac{v_\phi}{\Lambda}y_{\alpha 2},
		\label{Yukawa}
	\end{multlined}
	\eeq
	once the $U(1)_L$ symmetry is spontaneously broken. 
	
	\begin{figure}[!hbt]
		\centering
		\includegraphics[scale=0.55]{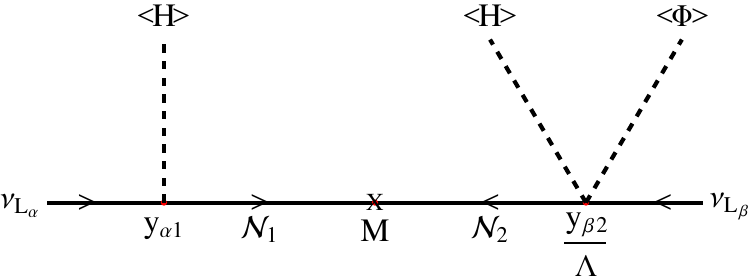}\\
		\caption{Schematic diagram for generation of the light neutrino mass, $m_\nu$.}
		\label{seesaw}
	\end{figure}
	
	As a result, after the EWSB (by then, $U(1)_L$ is already broken), the mass terms involving $\nu_L$ and $N_{R_i}$ $(i=1,2)$ can be cast (suppressing the generation indices) into 
	\beeq
	-\mathcal{L}_{\nu}\supset \frac{1}{2}\begin{pmatrix} \overline{\nu_L} & \overline{N_{R}^C}  \end{pmatrix}\begin{pmatrix}
		0 & m_D \\
		m_D^T & M_R^{d}
	\end{pmatrix} \begin{pmatrix} \nu_L^C \\  N_{R} \end{pmatrix}.
	\label{eq:nu-mass-matrix}
	\eeq
	The light and heavy neutrino mass matrices can now be obtained (with $m_D < M_R^d$) in the rotated basis $(\tilde{\nu}^c_L~N_R)^T$ via the type-I seesaw relations \cite{Ibarra:2003up}: $m_{\nu}\simeq m_D {(M_R)}^{-1} m_D^T$ and $M_R \equiv M_R^d$ (already diagonal), where $\tilde{\nu}^c_L=-i({\nu}^c_L-\theta_{\rm as} N_R)$ and $\theta_{\rm as}=m_D^* M_R^{-1}$ being the active-sterile neutrino mixing matrix. Finally, after $m_\nu$ is futher diagonalised by the PMNS matrix $(U)$ \cite{ParticleDataGroup:2020ssz}, final neutrino mass eigenstates $(n_j~N_{R_i})^T$ can be obtained. Hereafter, we shall use the notation of Majorana mass basis $\nu_j=n_j+n_j^c$ and $N_i=N_{R_i}+N_{R_i}^c$ for all the interactions. The RHN masses, written in this Majorana mass basis, and their mass-splitting ($\Delta M$) turn out to be\footnote{{With $\kappa_1 \neq \kappa_2$, mass splitting would have an additional parameter $(c_1 +  c_2)/2 \neq 1$, the effect of which is just to redefining/shifting the values of the parameters involved so as to remain consistent with the phenomenology of the scenario, without loss of generality.}}:
	\begin{align}
		M_{1,2}=\frac{fv_\phi}{\sqrt{2}} \pm \frac{v_\phi^2}{\Lambda}; ~~
		\Delta M=2\frac{v_\phi^2}{\Lambda}.
	\end{align}
	
	Note that the neutrino Yukawa coupling matrix $y^{\nu}$ plays an important role in fixing the neutrino mass, mixing as well as the amount of asymmetry in leptogenesis, which will be evident as we proceed. Here, it is then pertinent to discuss its expected order of magnitude from the neutrino oscillation data \cite{deSalas:2020pgw}, {considering a normal hierarchy of the light neutrinos. In addition, the presence of two RHNs in the present setup
		compels us to consider the lightest active neutrino mass to be zero.} The structure of $y^\nu$ can be estimated using the Casas Ibarra (CI) parametrization \cite{Casas:2001sr}
	\beeq
	y^\nu=\frac{\sqrt{2}}{v}U^{\dagger} D_{\sqrt{m_\nu}}R D_{\sqrt{M_R^d}},
	\eeq
	where $D_{\sqrt{m_\nu}}$ $(D_{\sqrt{M_R^d}})$ is the squared root of the $3\times3$ ($2\times2$) diagonal active neutrino \cite{Esteban:2016qun,Esteban:2020cvm} (RHN) mass matrix. $R$ is a complex orthogonal matrix, of the form \cite{Ibarra:2003up}
	\beeq
	R=\begin{pmatrix}
		0 & 0\\
		\textrm{cos}~\theta_R & 	\textrm{sin}~\theta_R\\
		-\textrm{sin}~\theta_R & 	\textrm{cos}~\theta_R
	\end{pmatrix},
	\eeq
	where $\theta_R$ is a complex angle in general $(\theta_R=z_R+iz_I)$. For concreteness, we provide a typical $y^\nu$ below 
	\beeq
	y^\nu=\begin{pmatrix}
		-0.0023-0.00043i & 0.0013-0.0011i \\
		0.0029+0.00053i & -0.0012+0.001i \\
		-0.0013-0.0014i & 0.0035-0.00050i \end{pmatrix},
	\label{yukawa}
	\eeq
	which is obtained using a set of reference values of the parameters (involved in parameter space scan of the upcoming sections), as shown in the following table \ref{t0} along with the choice of $\theta_R=\frac{\pi}{4}+0.42i$. Such a choice of parameters implies $M_1=1.4\times 10^{10}$ GeV and $M_2=1.36\times 10^{10}$ GeV.
	\begin{table}[hbt]
		\begin{center}
			\begin{tabular}{|c|c|c|}
				\hline
				$\Lambda$ (GeV) & $v_\phi$ (GeV) & $f$ \\
				\hline
				$1.5\times 10^{14} $  & $1.95\times 10^{11}$ &  $0.1$ \\                 
				\hline
			\end{tabular}
		\end{center}  
		\caption{{Reference values of the parameters used for $y^\nu$ estimation (see Eq. \ref{yukawa})}.}
		\label{t0}
	\end{table}
	It is also interesting to point out that such a $y^{\nu}$ is consistent with Eq. \ref{Yukawa} with the choice of $y_{\alpha 1} \sim \mathcal{O}(10^{-3})$ where $y_{\alpha 2} \sim \mathcal{O}(1)$ as $\frac{v_\phi}{\Lambda} = 10^{-3}$ is already of similar 
	order as $y^\nu$. 
	
	\section{DM phenomenology}
	\label{DM}
	
	We now turn our discussion on the Majoron being a dark matter candidate and for that purpose, we plan to estimate the number density of Majorons. Due to its pNGB nature, its interactions with the SM sector are suppressed by the $U(1)_L$ breaking scale, $v_\phi$, and hence automatically helps its stability. From the point of view of its production, we first show below that the tree-level coupling of Majorons to RHNs, via Eq. \ref{eq:seesaw}, is not sufficient to produce enough Majorons. Consequently, we find that the production of Majorons take place via a freeze-in scenario and we elaborate on such production channels and stability criteria of Majoron to be a successful dark matter candidate. 
	
	\subsection{Majoron production from RHNs {and stability}}
	The interaction between Majoron and RHNs as in Eq. \ref{eq:seesaw} turns out to be, 
	\beeq
	-\mathcal{L}\supset \frac{i f}{2\sqrt{2}}\chi N_i \gamma^5 N_i,
	\eeq
	after the $U(1)_L$ symmetry is spontaneously broken. This further initiates the following interaction involving Majoron, RHNs and light neutrinos once the EWSB takes place, 
	\beeq
	\begin{multlined}
		-\mathcal{L}_{\chi N \nu} =-\sum_{i,j}\mathcal{L}_{\chi N_i \nu_j} \\
		= \frac{\chi}{2\sqrt{2}}   \sum_{i,j} f_i \left(\overline{\nu_j} P_R N_i V^T_{ji}  + \overline{N_i}P_R \nu_j V_{ij}\right) + h.c.,
	\end{multlined}
	\label{eq:chi-N-nu}
	\eeq
	via the active sterile neutrino mixing $\theta_{\rm as}$ appearing in $V = \theta_{\rm as}^\dagger U$. The interaction of Eq. \ref{eq:chi-N-nu} results into a possible production channel of Majoron through the decay of RHNs $(N_i\to\chi\nu)$. The smallness of the active-sterile neutrino mixing indicates that this Majoron production may lead to a FIMP type dark matter \cite{Hall:2009bx} where the associated decay width of RHN is given by 
	\beeq
	\Gamma_{N_i\to\chi \nu}= \frac{M_i^3}{32 \pi v_\phi^2}  \sum_{j=1,2,3} |V_{ij}|^2.
	\label{d1}
	\eeq 
	
	Note that, there also exists a channel through which Majoron can decay into two light neutrinos, having the Lagrangian, 
	\beeq
	\begin{multlined}
		-\mathcal{L}_{\chi \nu \nu} =-\sum_{j,k}\mathcal{L}_{\chi \nu_j \nu_k}\\
		=-\frac{\chi }{2\sqrt{2}} \sum_{j,k} \left(\sum_{i}if_i \overline{\nu_j} P_R \nu_k V^T_{ji} V_{ik} +h.c. \right).
	\end{multlined}
	\label{eq:chi-nu-nu}
	\eeq
	and the corresponding decay width be given by 
	\beeq
	\Gamma_{\chi\to \nu \nu} = \frac{m_\chi}{16\pi v_\phi^2} \sum_{j} m_{\nu_j}^2.
	\label{majoron-instability}
	\eeq
	Hence, when the stability condition of Majoron dark matter with respect to the Universe lifetime ($\sim \mathcal{O}(10^{19})$ sec) is taken into account, the relic density of Majoron dark matter (produced via $N_i\to\chi\nu$) turns out to be negligible compared to present DM relic density (see \cite{Manna:2022gwn} for more details). Another interaction of Majoron could be present with the CP even scalar $\phi$. {However, we assume $\phi$ to be a heavy scalar $(m_\phi\sim v_\phi)$ and alongside that, with $\lambda_{H\Phi}$ assumed to be negligible, $\phi$ can be considered to be decoupled from the thermal bath.} The other possibility of $\phi$ mediating Majoron DM production from the annihilations of the Higgs $(HH\to\chi\chi)$ is mainly applicable to TeV scale Majoron \cite{Abe:2020dut}. In this work, our focus is on the lighter Majorons. 
	
	In this work, we introduce dimension-5 lepton number violating operators and show how these lead to a viable phenomenology of the Majoron as a freeze-in dark matter candidate, with the assumption that Majoron was initially absent in the early Universe. As mentioned earlier, the inclusion of the additional higher dimensional operators in Eq. \ref{eq:dim5-explicit} induce the following interactions after the $U(1)_L$ symmetry gets spontaneously broken as
	\beeq
	\begin{multlined}
		\frac{(\Phi^2+(\Phi^*)^2)\overline{\mathcal{N}_1^C}\mathcal{N}_1}{2\Lambda}+\frac{(\Phi^2+(\Phi^*)^2)\overline{\mathcal{N}_2^C}\mathcal{N}_2}{2\Lambda}\\
		\supset \frac{\chi^2}{2\Lambda} (\overline{N_1^C} N_1-\overline{N_2^C} N_2).
	\end{multlined}
	\label{eq:chi-production}
	\eeq 
	With such effective interactions, Majorons could be dominantly produced from the annihilations of RHNs, which leads to a signature of UV freeze-in \cite{Elahi:2014fsa,Barman:2020plp,Barman:2021tgt,King:2023ztb} as we discuss below. {Note that any deviation from the universal coupling (ensured by assuming the CP symmetry, $\Phi \rightarrow \Phi^*$) with $\Phi$ and $\Phi^*$ of the Yukawa interaction may result Majoron instability (contributing toward $\chi \rightarrow \nu \nu$) due to the presence of higher dimensional terms in the construction (see appendix \ref{apndx_A}). However, such decay widths being suppressed (or at most of similar order) compared to the one in Eq. \ref{majoron-instability}, the inclusion of Majoron's lifetime greater than the age of the Universe would automatically take care of this constraint, making it a viable dark matter candidate in our construction.}

	\subsection{Majorons as Freeze-in dark matter}
	
	{With the effective interactions of Majorons with RHNs as indicated in Eq. \ref{eq:chi-production}, the Majorons can be produced from the annihilations of RHNs in the early Universe. In this case, contrary to the usual freeze-in scenario of Majoron production from RHN decay,} the production of Majoron dark matter crucially depends on the maximum temperature of the thermal bath. Throughout our analysis, {we shall consider 
		the breaking of $U(1)_L$ symmetry to take place prior to the reheating temperature $(T_{\rm RH})$ of the Universe}, $i.e$ during inflation or in the reheating phase, with $v_\phi >T_{\rm RH}$. Such a choice of hierarchy between $v_{\phi}$ and $T_{\rm RH}$ is also consistent with our assumption of $\phi$ being decoupled from the study below $T_{\rm RH}$. Hence, in our case, dark matter production is sensitive to $T_{\rm RH}$ which is considered as a free parameter in our study. {We further consider the RHNs to be present in the thermal bath requiring $T_{\rm RH}>M_i$ and a sizeable neutrino Yukawa interactions so as they can be responsible for thermal leptogenesis}. As a result, a specific hierarchy between the energy scales $v_\phi>T_{\rm RH}>M_i$ is essential for our setup. 
	{Furthermore, since both the $U(1)_L$ and $Z_2$ symmetries are spontaneously broken when $\Phi$ obtains a non-zero $vev$ prior to $T_{\rm RH}$ here ($i.e.$ during inflation or extended reheating period), any topological defects formed during this symmetry breaking would be diluted during inflation.}
	
	To analyse the evolution of DM yield originated from the annihilations of RHNs as discussed above, we solve the following Boltzmann equation, expressed in terms of the co-moving number density $Y_\chi~(=n_\chi/\mathcal{S})$ and the temperature $(T)$ as follows \cite{Kolb:1990vq,Chianese:2019epo}
	\beeq
	\begin{multlined}
		\frac{dY_\chi}{dT} \simeq  -\frac{2\mathcal{S}}{\mathcal{H}T} \sum_{i} \big\langle\sigma v\big\rangle_{N_iN_i\to\chi\chi} Y_{N_{i}}^{{\rm eq}^2} \left[ 1-\frac{Y_\chi^2}{Y_{\chi}^{{\rm eq}^2}}\right], 
		\label{BEbasic}
	\end{multlined}
	\eeq
	where $\mathcal{S}=\frac{2\pi^2}{45}g_\star^\mathcal{S}(T)T^3$ and $\mathcal{H}=1.66\sqrt{g_\star^\rho(T)}\frac{T^2}{M_{\rm Pl}}$ denote the entropy density and the Hubble expansion rate of the Universe, respectively. Here, $M_{\rm Pl}=1.22\times 10^{19}$ GeV represents the Planck scale and $g_\star^\mathcal{S}(T)$ and $g_\star^\rho(T)$ describe the effective number of relativistic degrees of freedom at temperature $T$ ($e.g.$ at high temperature, $g_\star^\mathcal{S}(T)=g_\star^\rho(T)=106.75$, constituting out of the SM particle content). $Y^{{\rm eq}}$ in Eq. \ref{BEbasic} denotes the equilibrium yield of a particle species (say $X$) as
	\beeq
	Y_{X}^{{\rm eq}} (T)= \frac{45 g_X}{4\pi^4 g^\mathcal{S}_{\star}(T)}\left( \frac{m_X}{T}\right)^2 K_2\left(\frac{m_X}{T} \right),
	\eeq
	where $m_X$ and $g_X=1(2)$ (for scalar (fermion)) indicate the mass and the internal degrees of freedom of the particle species $X$, respectively, and $K_2$ is the modified Bessel's function of second kind. In Eq. \ref{BEbasic}, another production channel of Majoron $e.g.$ $N_i\to\chi\nu$ is not included as it contributes negligibly to the DM abundance being suppressed by the active-sterile neutrino mixing. Another contribution to the process $N_iN_i\to\chi\chi$ proceeds via $t$-channel RHN mediation, which however drives IR freeze-in \cite{Hall:2009bx} and remains subdominant compared to our scenario as discussed in \cite{Frigerio:2011in,Abe:2020dut}.  Here, $\big\langle\sigma v\big\rangle_{N_iN_i\to\chi\chi}$ is the thermally averaged cross-section of the process $N_iN_i\to\chi\chi$, expressed as
	\beeq
	\begin{multlined}
		\big\langle\sigma v\big\rangle_{N_iN_i\to\chi\chi}=
		\frac{1}{8M_i^4TK_2^2(M_i/T)}\times\\
		\int_{4M_i^2}^{\infty}\sigma_{N_iN_i\to\chi\chi} (s-4M_i^2)\sqrt{s}K_1\left( { \sqrt{s}}/{T} \right)  ds,
	\end{multlined}
	\label{sigmav}
	\eeq
	where $\sigma$ denotes the cross-section of the process ${N_{i}N_{i}\to \chi\chi}$ as
	\beeq
	\sigma_{N_{i}N_{i}\to \chi\chi}=\frac{1}{16\pi s}\sqrt{\frac{s-4m_{\chi}^2}{s-4M_{{i}}^2}} |\overline{M}|^2_{N_{i}N_{i}\to \chi\chi},
	\label{cross-sec}
	\eeq
	with 
	\beeq
	|\overline{M}|^2_{N_{i}N_{i}\to \chi\chi}= \frac{1 }{\Lambda^2}\left(s-4M_i^2 \right).
	\label{ampl}
	\eeq
	The final yield of the Majoron DM can be obtained by integrating Eq. (\ref{BEbasic}) from $T_{\rm RH}$ (reheating temperature) to $T_0$ (present temperature). One then obtain the final relic density of $\chi$ by replacing the present DM abundance $Y_\chi(T_0)$ in
	\beeq
	\Omega_\chi h^2=2.755\times10^8\left(\frac{m_\chi}{\rm{GeV}} \right)Y_\chi(T_0),
	\eeq  
	{which should reproduce the observed dark matter relic density, $\Omega_{\rm DM}h^2=0.12\pm 0.0012$ \cite{Planck:2018vyg}}.

	The main parameters involved in our analysis of dark matter production are $f,~\Lambda,~T_{\rm RH} ~\&~m_\chi$. Another important parameter is $v_\phi$ which enters in the RHN masses as $M_i=\left(\frac{fv_\phi}{\sqrt{2}}\pm \frac{v_\phi^2}{\Lambda} \right) $. However, for the DM phenomenology, the role of $v_\phi$ is not very significant as the DM yield, being dominated by UV freeze-in scenario, primarily depends on $\Lambda$ and $T_{\rm RH}$ rather than the masses of the parent particles. Specifically, the Majoron yield $(Y_\chi)$ approximately follows the relation $Y_\chi\propto T_{\rm RH}/\Lambda^2$ as evident from Eqs. (\ref{BEbasic}) - (\ref{ampl}).

	\begin{figure}[!h]
		\centering
		\includegraphics[scale=0.5]{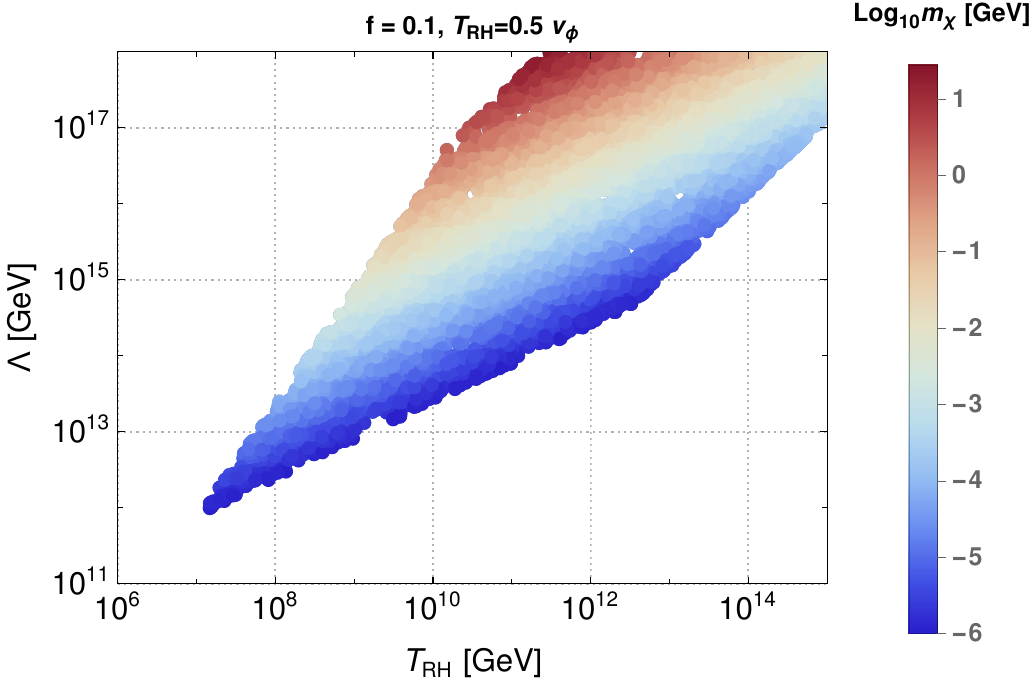}\\
		\includegraphics[scale=0.55]{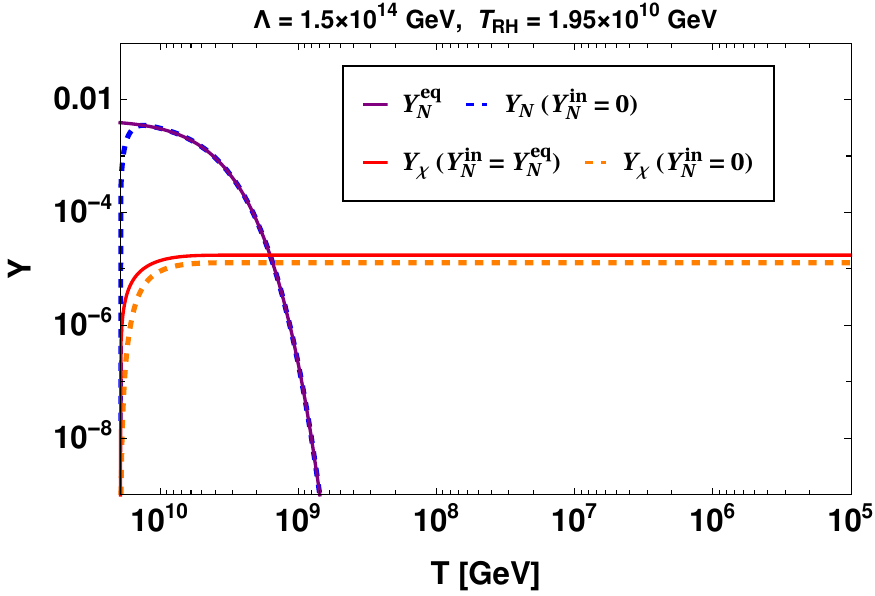}
		\caption{Relic satisfied parameter space represented by colored region in $T_{\rm RH}$ vs $\Lambda$ plane, while $m_\chi$ are in color side-bar (upper) and a benchmark point (lower) are shown. Here $f=0.1$ and the ratio $T_{\rm RH}/v_\phi=0.5$ is considered.}
		\label{Fig-DM}
	\end{figure}
	
	{Since $v_{\phi}$ is not directly entering in the DM phenomenology except its involvement in the RHN mass and splitting (which are anyway insensitive to Majoron yield), we have chosen certain (mild) hierarchies between $v_\phi$ and $T_{\rm RH}$ as benchmark values for our analysis. For example, in Fig. \ref{Fig-DM}, a ratio of $T_{\rm RH}/v_\phi=0.5$ and $f=0.1$ is chosen.} The upper panel of Fig. \ref{Fig-DM} illustrates the DM parameter space in the $T_{\rm RH}-\Lambda$ plane with $m_\chi$, obtained in order to satisfy the correct relic for each point in the parameter space, is shown in the color bar (with blue to dark red color gradient). As can be seen from  Fig. \ref{Fig-DM} (upper panel), we scan over a large region of the parameters; $
	(i)~T_{\rm RH}: 10^6~ {\rm{GeV}} - 10^{15} ~{\rm{GeV}} ;
	(ii)~\Lambda: 10^{11} ~{\rm{GeV}} - 10^{18}~ {\rm{GeV}}~\&~
	{(iii)~m_{\chi} \gtrsim 1~ {\rm{keV}}.}
	\nonumber
	$ While doing the scan, the conditions imposed on the parameters are the following: (a) $v_\phi>T_{\rm RH}>M_i$ as stated earlier and (b) $f/\sqrt{2}>v_\phi/\Lambda$ which follows from the fact that the contribution to the RHNs mass from the dimension-5 operator must remain subdominant compared to the renormalizable one. While {boundary at the right side stems from the condition $f/\sqrt{2}>v_\phi/\Lambda$ (considering the ratio of $T_{\rm RH}/v_\phi=0.5$), the left-side boundary signifies the Majoron DM must be stable following the essential criteria, $\Gamma_{\chi\to\nu\nu}^{-1}>\tau_U$. On the other hand, the
		top and the bottom boundary of the parameter space signifies the viable Majoron DM mass range, $i.e.$ $\mathcal{O}(100)$ GeV $\geq m_\chi\geq$ 1 keV, as evident from the color bar}.  In the lower panel of Fig. \ref{Fig-DM}, we have shown the DM yield ({solid red line}) for the benchmark point mentioned in table \ref{t0}, which falls within the allowed parameter space of the upper panel. As expected, the maximum production of the DM occurs at the very initial temperature, which is $T_{\rm RH}$. 
	\begin{figure}[!hbt]
		\centering
		\includegraphics[scale=0.5]{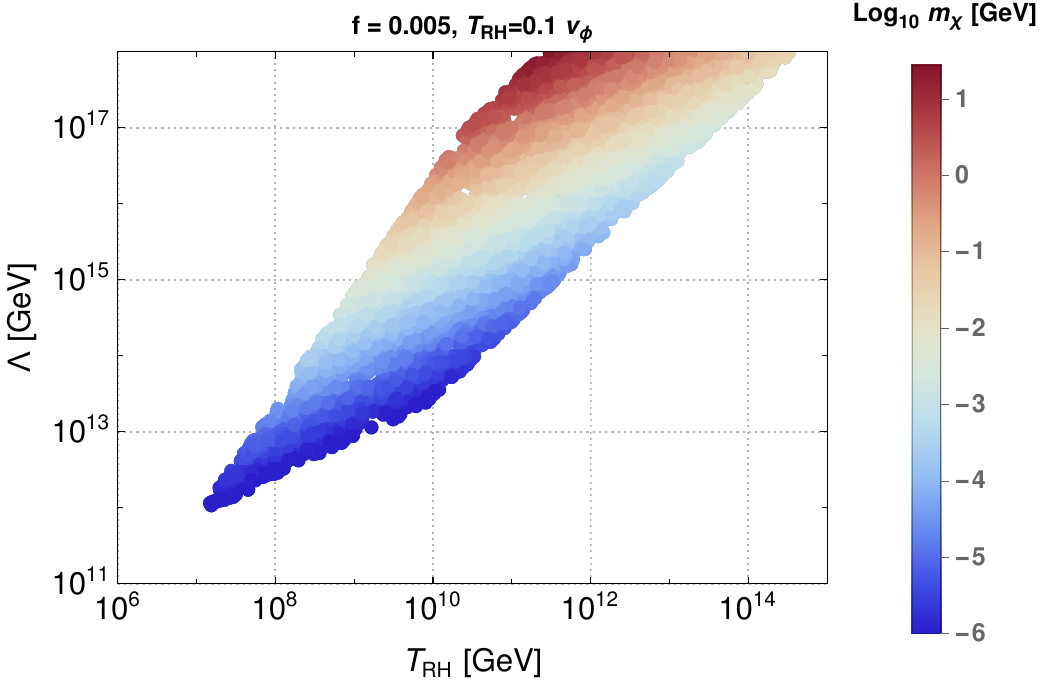}\\
		\caption{Relic satisfied parameter space represented by colored region in $T_{\rm RH}$ vs $\Lambda$ plane, while $m_\chi$ are in color side-bar with $f=0.005$ and the ratio $T_{\rm RH}/v_\phi=0.1$.}
		\label{Fig-DM2}
	\end{figure}
	In Fig. \ref{Fig-DM2}, we have shown the relic satisfied parameter space with a difference choice of ratio $T_{\rm RH}/v_\phi=0.1$. It turns out that the parameter space in this case (in Fig. \ref{Fig-DM2}) gets shrunk compared to Fig. \ref{Fig-DM} due to less number of points, satisfying the criteria $v_\phi>T_{\rm RH}>M_i$ and $f/\sqrt{2}>v_\phi/\Lambda$.

	{It is perhaps pertinent to discuss the situation, contrary to our earlier assumption, where the RHNs are not part of the thermal bath initially at $T_{\rm RH}$. In this case, the RHNs would attain thermal equilibrium through the inverse decay $LH\to N_i$ process, proportional to the neutrino Yukawa coupling $y^\nu_{\alpha i}$. However, how quickly they attain equilibrium depends solely on the strength of $y^\nu_{\alpha i}$. Since the Majorons are mainly produced from the annihilation of these RHNs in our framework, such Majoron production could be delayed depending on the time the RHNs would take to thermalise themselves. In order to explore that, we plot the yield of the RHNs for the same set of parameters ($\Lambda, f, T_{\rm RH}$) as used to plot $Y_N^{\rm eq}$ in Fig. \ref{Fig-DM}, but with zero initial abundance now, and depict it by {blue dashed line}. As seen, $Y_N$ reaches the equilibrium abundance at a temperature slightly below $T_{RH}$ due to the presence of the sizeable neutrino Yukawa coupling $y^\nu_{\alpha i}\sim \mathcal{O}(10^{-3})$, corresponding to the benchmark values of Tab. \ref{t0}. This tiny delay in reaching thermal equilibrium for the RHNs results a relatively less freeze-in abundance of Majorons ({orange dashed line}). Hence, the correct relic abundance for the Majoron as DM would follow with a mild shift in $m_{\chi}$ value (from earlier value $2.5 \times 10^{-5}$ GeV to a new one, 3.5$\times 10^{-5}$ GeV). Since the neutrino Yukawa remains sizeable enough for the entire parameter space, our conclusion with thermalised RHNs at $T_{\rm RH}$ would mostly remain unchanged in case of zero abundance of RHNs, apart from a slight modification in Majoron mass to a relatively higher side.}

	{In addition to the Majoron DM stability condition, another stringent constraint follows from monochromatic neutrino searches by experiments like Borexino \cite{Borexino:2010zht}, KamLAND \cite{KamLAND:2011bnd}, SuperKamiokande {\cite{Super-Kamiokande:2002exp,Super-Kamiokande:2013ufi}}, and IceCube {\cite{IceCube:2011kcp}}, due to the model-independent decay of Majorons into light neutrinos. These experiments restricts the Majoron parameter space for $m_\chi\gtrsim 4$ MeV \cite{Heeck:2017xbu,Akita:2023qiz}. Additionally, Majoron decay into two photons (induced in two-loops) provides another channel for probing the parameter space through several $\gamma$-ray observations \cite{Heeck:2017xbu,Heeck:2019guh}. We have discussed these constraints in detail in an earlier work \cite{Manna:2022gwn}. }
	
	\section{Resonant Leptogenesis}
	\label{leptogenesis}
	
	After getting the DM allowed parameter space, we are now in a position to discuss the leptogenesis scenario keeping in mind that the same set of operators of Eq. \ref{eq:chi-production} contributing toward Majoron production will also be responsible for breaking the degeneracy of RHN mass, hence playing a key role in leptogenesis too. Before entering the details of the leptogenesis, specific to our set-up, we provide a brief overview of leptogenesis in general, more specifically the resonant case. The dynamical generation of lepton asymmetry takes place in an the era of radiation dominated Universe, below\footnote{For a generation of lepton asymmetry during the reheating phase $i.e.$ when $M_i > T_{\rm RH}$, we refer the readers to \cite{Lazarides:1991wu,Murayama:1992ua,Kolb:1996jt,Giudice:1999fb,Asaka:1999yd,Asaka:1999jb,Hamaguchi:2001gw,Barman:2021ost,Barman:2022gjo,Lazarides:2022ezc,Barman:2024jqh,Barman:2024ujh}.} $T_{\rm RH}$ {($M_i < T_{\rm RH}$)}, when the RHNs  starts decaying. 
	
	Below the reheating temperature, the RHNs can be produced from the thermal bath due to inverse decay via the neutrino Yukawa interaction. The same interaction also keeps them in the equilibrium with the SM bath till the temperature of the Universe remain larger than their masses. Thereafter, their out of equilibrium decay to the SM 
	Higgs and lepton doublet begins having a decay width, given by
	\begin{equation}
		\Gamma_{N_i\to LH}=\frac{|y_{ii}^\nu|^2}{8\pi} M_i.
	\end{equation}
	One can track the abundance of the RHNs in the expanding Universe by solving the following Boltzmann equations:
	\bea
	\frac{dY_{N_i}}{dx}&=& -\frac{1}{\mathcal{H}x\mathcal{S}} \left[\frac{Y_{N_i}}{Y_{N_i}^{eq}}-1\right](\gamma_{N_i}+2\gamma_{h_s}+4\gamma_{h_t}),\nonumber\\
	\label{BE_RHN}
	\eea
	where \beeq \gamma_{N_i}=n_{N_i}^{eq}\frac{K_1(x)}{K_2(x)}\Gamma_{N_i},
	\eeq
	$x=M_1/T$ with $T$ being the temperature of the Universe. Additionally, $\gamma_{h_s}$ and $\gamma_{h_t}$ signify the reaction rates of Higgs mediated $\Delta L=1$ processes involving the SM top quarks, $e.g.,$ $N_i L \to qt$ (s-channel) and $N_i t \to Lq$ (t-channel), which are defined by
	\beeq
	\gamma_{h_{s,t}}=\frac{M_i}{64\pi^2 x}\int ds \hat{\sigma_h} (s)\sqrt{s} K_1\left(\frac{x\sqrt{s}}{M_i} \right),
	\label{DL=1}
	\eeq
	where
	\beeq
	\hat{\sigma_h} (s) =\frac{y_t^2(y^{\nu^\dagger}y^\nu)_{ii}}{4\pi}\left(1-\frac{M_i^2}{s} \right)^2. 
	\eeq
	Finally, $Y_{N_i}=n_{N_i}/\mathcal{S}$ denotes the comoving number density of the RHNs with $n_{N_i}$ being the number density of the RHNs. Looking at Eq. \ref{BE_RHN}, one notices that the first term in the squared bracket on the right hand side (r.h.s) comes with a negative sign and hence is responsible for the depletion of RHNs' abundance resulting {primarily} from its decay to the SM particle, while the second term is responsible for their production from bath due to the inverse decay. 
	
	The out of equilibrium decays of RHNs together with the lepton number violation (present due to Majorana masses of RHNs) and the CP violation originating from the neutrino Yukawa sector (as evident from the structure of $y^{\nu}$ via CI parametrisation) are the three necessary and sufficient conditions, namely the Sakharov's conditions \cite{Sakharov:1967dj}, required for the dynamical generation of lepton asymmetry. It can be inferred from the DM phenomenology of the previous section that the RHN masses in our case ($\sim f v_{\phi}$), satisfying the correct relic, fall in a broad range from a very heavy to as light as {$10^4$} GeV. In this connection, one can recall that the standard thermal leptogenesis works for the RHN mass above $10^9$ GeV, the so-called Davidson-Ibarra bound \cite{Davidson:2002qv}. However, this conclusion is based on the hierarchical nature of RHNs. In case degenerate RHNs are present, like the present case, the asymmetry production can be significantly enhanced. In order to quantify such enhanced production of asymmetry, we proceed below for the evaluation of the CP asymmetry. 
	
	The CP asymmetry parameter $\epsilon_{N_i}$ associated to the decay of the $i$-th RHN can be expressed as 
	\beeq
	\epsilon_{N_i} = \frac{\sum_{\alpha}[\Gamma_{N_i\to L_{\alpha} H}-\Gamma_{N_i\to \bar{L}_{\alpha} \bar{H}}]}{\sum_{\alpha}[\Gamma_{N_i\to L_{\alpha} H}+\Gamma_{N_i\to \bar{L}_{\alpha} \bar{H}}]},
	\eeq
	which results from the interference of the tree-level decay of $N_i$ and the one loop, vertex and self energy, diagrams. The general expression for such CP asymmetry {(after flavor sum)} can be estimated as \cite{Flanz:1996fb,Pilaftsis:1997jf,Pilaftsis:2003gt,Iso:2010mv,Qi:2022fzs,Das:2024gua}
	\beeq
	\epsilon_{N_i}=-\sum_{j\neq i}\frac{M_i \Gamma_{N_j}}{M_j^2}\left(\frac{V_{ij}}{2}+S_{ij} \right) \frac{{\rm Im}(y^{\nu^\dagger}y^\nu)^2_{ij}}{(y^{\nu^\dagger}y^\nu)_{ii}(y^{\nu^\dagger}y^\nu)_{jj}},
	\label{CP_asy}
	\eeq
	where
	\beeq
	V_{ij}=2\frac{M_j^2}{M_i^2}\left[\left(1+ \frac{M_j^2}{M_i^2}\right) {\rm ln}\left(1+ \frac{M_i^2}{M_j^2} \right)-1  \right],
	\eeq
	\beeq
	{\rm and} ~~~S_{ij}=\frac{M_j^2(M_j^2-M_i^2)}{(M_j^2-M_i^2)^2+M_i^2\Gamma_{N_j}^2},
	\eeq
	denote the vertex correction and self energy corrections respectively. At this stage, it is important to point out that though the contributions of the vertex and self energy corrections are of similar order for hierarchical RHNs, the self energy contributions dominate over the other in case of (closely) degenerate RHNs. This is mainly because, in the limit of quasi-degenerate RHNs masses $i.e.$ $M_i\approx M_j$, the CP asymmetry can be maximised with $M_i^2-M_j^2 \sim M_i \Gamma_{N_j}$ such that $S_{ij} \gg V_{ij} \sim \mathcal{O}(1)$ \cite{Pilaftsis:2003gt}. {Alongside, the remaining part of Eq. \ref{CP_asy} can be put in the form (with $\theta_R= z_R+i z_I$) \cite{Das:2024gua} 
		\beeq
		\begin{multlined}
			F_y = \frac{{\rm Im}(y^{\nu^\dagger}y^\nu)^2_{ij}}{(y^{\nu^\dagger}y^\nu)_{ii}(y^{\nu^\dagger}y^\nu)_{jj}}\\
			\approx \left| \frac{2(m_2^2-m_3^2)~{\rm sin} (2 z_R)~ {\rm sinh} (2 z_I)}{(m_2-m_3)^2 ~{\rm cos} (2 z_R)^2 - (m_2+m_3)^2 ~{\rm cosh} (2 z_I)^2}\right|,
			\label{eps-y}
		\end{multlined}
		\eeq
		\noindent	for normal hierarchical neutrinos which turns out to be $\mathcal{O}(0.7)$ for the choice of $\theta_R = 0.78+0.42i$ which we considered throughout the analysis and using the best fit values of neutrino mass splittings (with lightest neutrino eigenvalue $m_1$ set to zero) \cite{Esteban:2020cvm}. Note that this particular choice of 
		$\theta_R$ maximises the CP asymmetry, hence implying that 
		$F_y \lesssim \mathcal{O}(0.7)$ for other choices of $\theta_R$. Such an enhancement of the CP asymmetry} is the key ingredient of the resonant leptogenesis mechanism for which, even with light RHNs as low as TeV or so, it can satisfy the baryon asymmetry of the Universe via leptogenesis at the cost of imposing a precise degree of degeneracy between the two RHNs. In the present scenario of the minimal Majoron model, the splitting $\Delta M = 2 v_{\phi}^2/\Lambda$ is fixed for any specific point on the DM relic satisfied parameter space, as in Fig. \ref{Fig-DM} (upper panel) and \ref{Fig-DM2}. Therefore, for each such point, one can check the validity of the resonant leptogenesis mechanism in order to satisfy the BAU where the complex angle in the orthogonal $R$ matrix play crucial role. 
	
\begin{figure}[!bt]
		\centering
		\includegraphics[scale=0.5]{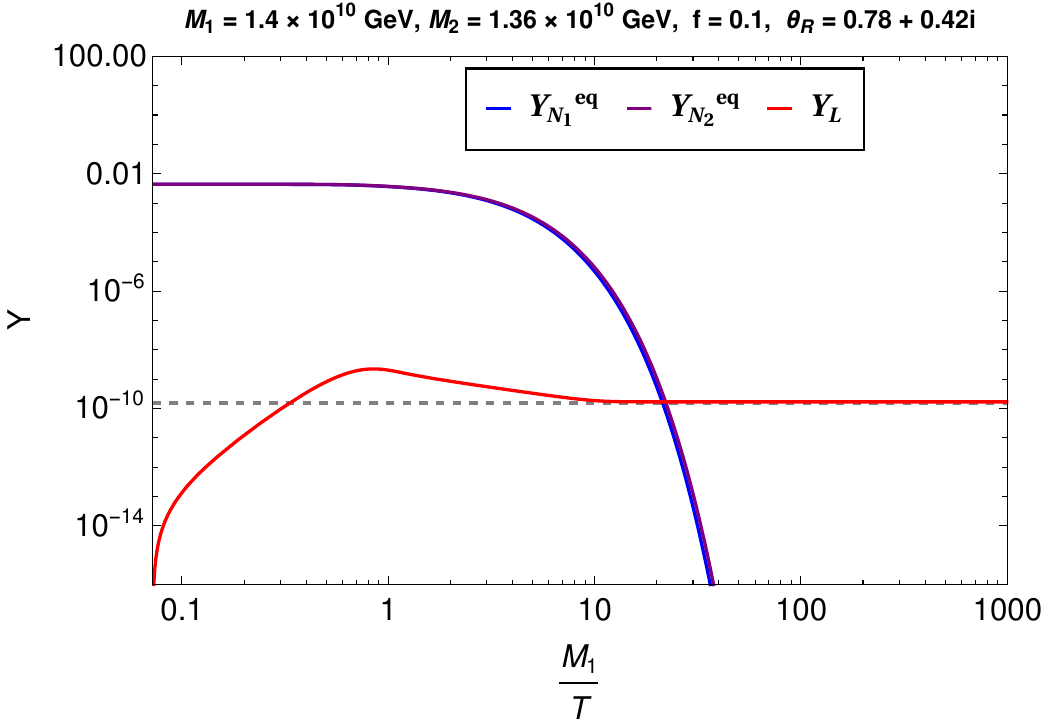}
		\caption{Evolution of the yields of RHNs and lepton asymmetry (using table \ref{t0}).}
		\label{BP-lep}
	\end{figure}

	Once the CP asymmetry is obtained, the evolution of the $L$ asymmetry can be studied by simultaneously solving the Boltzmann equation shown in Eq. \ref{BE_RHN}, together with \cite{Pilaftsis:1997jf,Plumacher:1996kc},
	\beeq
	\frac{dY_{L}}{dx}= \sum_{i}\frac{1}{\mathcal{H}x\mathcal{S}} \left[\epsilon_{N_i}\left(\frac{Y_{N_i}}{Y_{N_i}^{eq}}-1 \right)-\frac{Y_{L}}{2Y_{l}^{eq}}\right]\gamma_{N_i}\\-\frac{Y_L}{Y_{l}^{eq}}\gamma_{\sigma}
	\label{BE_asy}
	\eeq
	where $Y_L$ denotes the amount of asymmetry generated in the lepton sector and $\gamma_{\sigma}$ signifies the rates of $\Delta L=1$ scatterings processes, as mentioned in Eq. \ref{DL=1}. Here, the term proportional to $\epsilon_{N_i}$ in the r.h.s is responsible for the growth in the asymmetry $Y_L=Y_l-Y_{\bar{l}}$, which then gets washed out primarily due to the inverse-decays $LH\to N_i$ and $\bar{L}\bar{H}\to N_i$. Finally, as the temperature drops below $M_i$, the inverse decay processes get suppressed by $e^{-M_i/T}$, resulting in a saturation in the lepton asymmetry. Additionally, $\Delta L=2$ scattering processes such as $LH\to \bar{L}\bar{H}$ and $LL\to HH$ (mediated by $N_i$) also contribute in wash-out of the lepton asymmetry, which is however less efficient compared to the inverse decay processes, due to the involvement of $({y^\nu}^\dagger y^\nu)^2$. The asymptotic yield in lepton asymmetry $Y_L^{\infty}$ (at $x\to \infty$) is eventually converted into the baryon asymmetry $Y_B$ through electroweak sphalerons at temperatures above $T\sim 130$ GeV, as described by the relation {$Y_B=(28/51)Y_L^{\infty}\simeq 8.75\times 10^{-11}$} \cite{Davidson:2008bu}. {Here we do not incorporate the flavor effects on leptogenesis\footnote{{For flavor effects on thermal leptogenesis, we direct the readers to \cite{Abada:2006fw,Nardi:2006fx,Blanchet:2006be,Dev:2017trv,Datta:2021elq,Datta:2021gyi} and for flavor effects on leptogenesis during a non-instantaneous reheating epoch, we refer the readers to \cite{Datta:2022jic,Datta:2023pav}.}} for simplicity.} We use Fig. \ref{BP-lep} to demonstrate the evolution of the abundance of RHN together with the lepton asymmetry, $Y_L$ for the benchmark point shown in table \ref{t0} for which the observed baryon asymmetry is satisfied with $\theta_R = 0.78 + 0.42 i$.
	
	Notice that unlike the usual motivation of lowering the scale of RHNs (typically of order TeV or so) from its collider search point of view while employing the resonant leptogenesis scenario, the present construction requires a relatively heavier RHNs ($10^6$ GeV and above) to have a successful leptogenesis. This is because the DM phenomenology in our scenario demands a heavier RHN ($v_{\phi}>T_{\rm RH}$) together with larger $\Lambda$ as seen from Figs. \ref{Fig-DM} (upper panel) and \ref{Fig-DM2}. It is however ensured that the mass difference between the two almost degenerate RHNs are large enough compared to their respective decay width(s) so as to maintain the validity of the perturbative calculation. 
	As stated earlier, the mass splitting between the two RHNs depends on $v_{\phi}$ and $\Lambda$ and hence, is fixed for a specific set of values of $(v_{\phi}, \Lambda)$. It turns out that {$\epsilon_{N_i}\sim 6\times 10^{-6}$} is required to produce the lepton asymmetry that can explain the observed baryon asymmetry.
	
	\section{Correlating DM parameter space with Leptogenesis}
	\label{parameterspace}

	As discussed separately in the previous two sections, in this work, the explicit breaking of the lepton number at dimension 5 contributes to both the DM phenomenology and the generation of baryon asymmetry which in turn provides an interesting correlation between the DM parameter space to the generation of lepton asymmetry.  To demonstrate it, we consider the entire DM relic satisfied parameter space (as in upper panel of Fig. \ref{Fig-DM} and Fig. \ref{Fig-DM2}) and evaluate the corresponding baryon asymmetry of the Universe where the only additional parameter (apart from the common parameters: $\Lambda, ~T_{\rm RH}$ and $f$), entering solely in leptogenesis, is the complex angle $\theta_R$. Although the $U(1)_L$ breaking scale, $v_\phi$, is not directly involved in the DM analysis, it remains a crucial parameter for resonant leptogenesis as it directly contributes to the generation of tree level RHN (degenerate) masses and their mass-splittings. As mentioned in the DM phenomenological discussion, we assume $v_\phi>T_{\rm RH}$ and consider a fixed hierarchy between $T_{\rm RH}$ and $v_\phi$ for convenience. Specifically, we consider two benchmark hierarchies as (a) $T_{\rm RH}=0.5~ v_\phi$ and $f=0.1$ and (b) $T_{\rm RH}=0.1~ v_\phi$ for $f=0.005$. In addition, the ratio $T_{\rm RH}/v_\phi$ should satisfy $T_{\rm RH}/v_\phi\gtrsim f/\sqrt{2}$ which arises from the consideration, $T_{\rm RH}>M_i ~(\sim fv_\phi/\sqrt{2})$ ensuring that the RHNs can be considered to be in thermal equilibrium. {With these considerations, we scan the entire parameter space by varying $\Lambda, { T_{\rm RH}}$ and $m_{\chi}$ and look for the common parameter space satisfying the correct baryon asymmetry via leptogenesis and Majoron dark matter, as explained below. The impact of choosing different $\theta_R$ on such common parameter space is also highlighted.}
	
	\begin{figure}[t]
		\centering
		\includegraphics[scale=0.5]{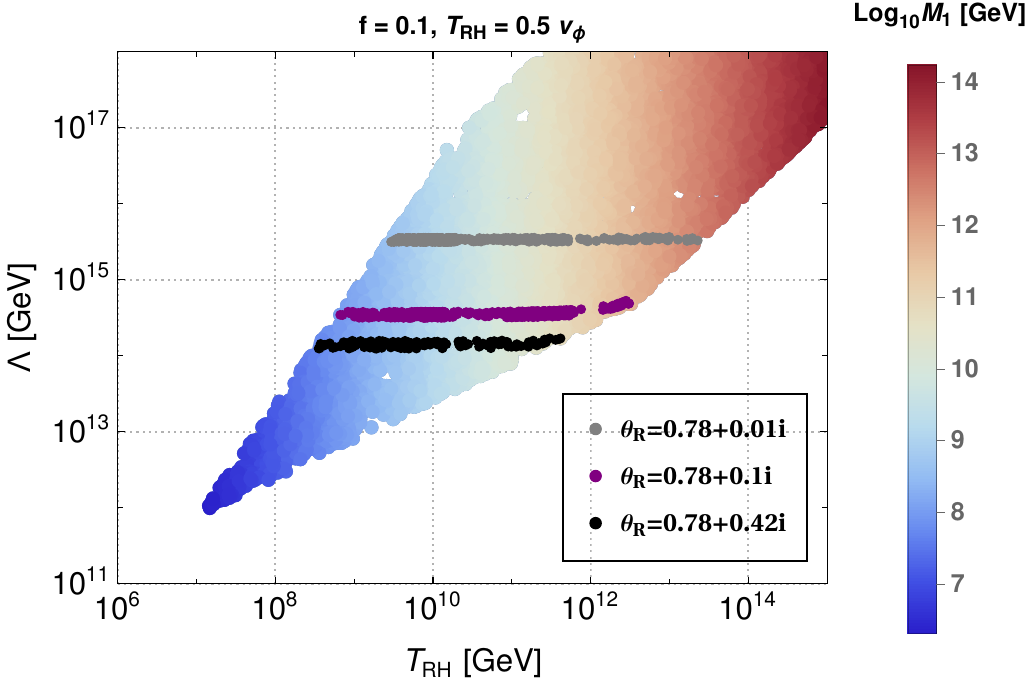} 
		\includegraphics[scale=0.5]{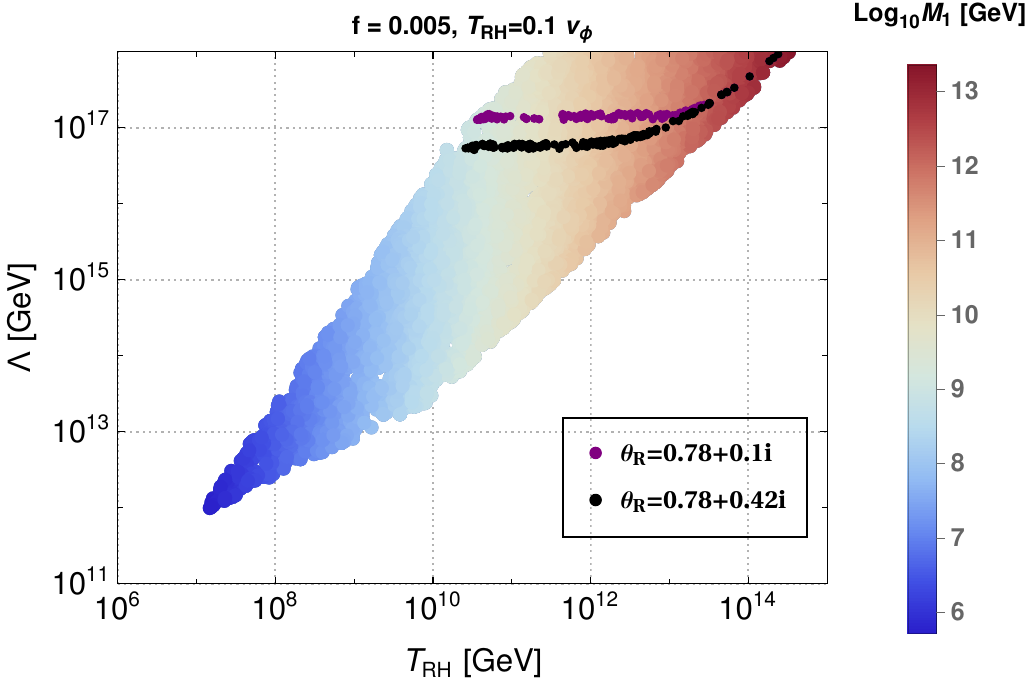}
		\caption{{The correct BAU satisfied parameter spaces are indicated on DM-parameter space (from upper panel of Fig. \ref{Fig-DM} and Fig. \ref{Fig-DM2}) by black, purple and gray patches, for different choices of $\theta_R$ (indicated in plot legends).}}
		\label{fig:common-ps}
	\end{figure}
	
\begin{figure}[tb]
		\centering
		\includegraphics[scale=0.5]{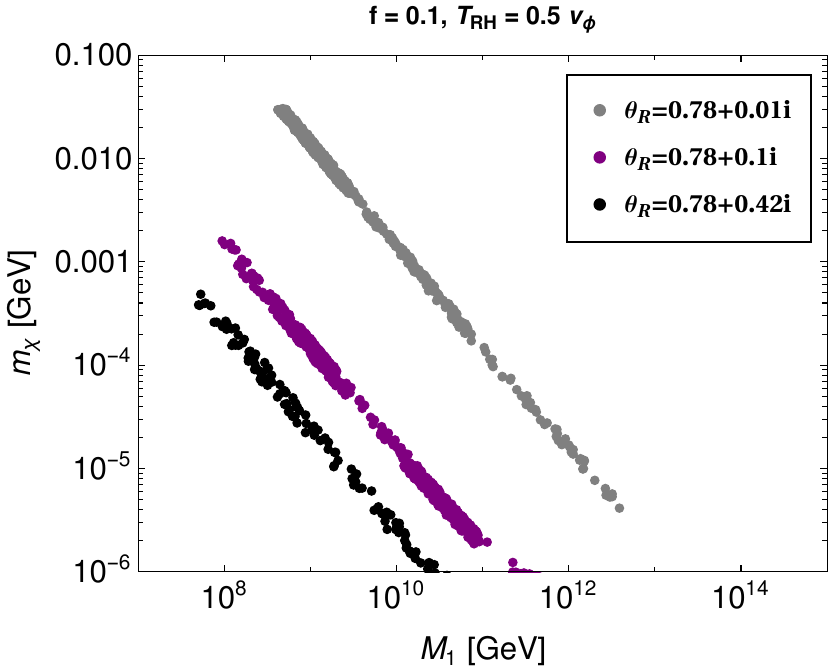}\\ 
		\includegraphics[scale=0.5]{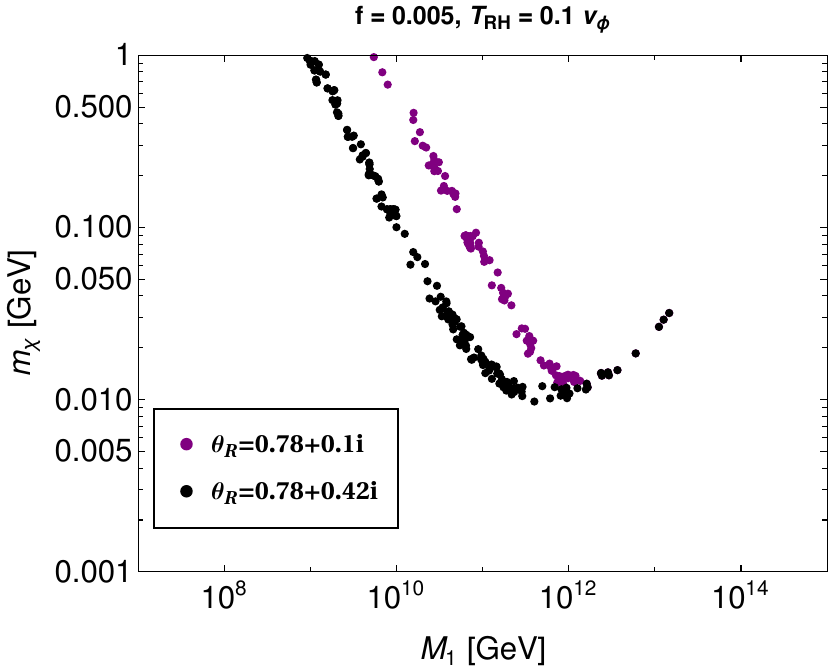}
		\caption{Correlation between masses of Majoron and RHNs, obtained from DM-relic and BAU satisfaction.}
		\label{fig:MN-mchi}
	\end{figure}

	In Fig. \ref{fig:common-ps}, we illustrate a common parameter space for Majoron as DM and resonant leptogenesis signifying the correct baryon asymmetry of the Universe in a single comprehensive plot in $\Lambda - T_{\rm RH}$ plane. {Additionally, the variation of the RHN mass is shown in color bar.}
	In the upper (lower) panel of Fig. \ref{fig:common-ps}, {the baryon asymmetry and DM relic satisfied parameter space is indicated by the black patch with $T_{\rm RH}/v_\phi =0.5 ~(0.1)$ and $f=0.1 ~(0.005)$ for 
		an optimum choice of $\theta_R = 0.78 + 0.42i$. It is observed that this particular $\theta_R$ maximizes the CP asymmetry parameter.} 
	It then turns out that for such choices of $f$ and $T_{\rm RH}/v_\phi$ in the upper plot of Fig. \ref{fig:common-ps}, successful resonant leptogenesis is obtained for RHN masses in the range $10^{8}-10^{11}$ GeV, while the Majoron as DM (satisfying correct relic density) falls in the mass range, $m_\chi \sim \mathcal{O}({\rm keV - MeV})$, {this is also visible from Fig. \ref{fig:MN-mchi}}. 
	Similarly, for a lower choice of $f~(=0.005)$ (associated to smaller RHN masses), the mass splitting must be more fine-tuned, demanding higher values of $\Lambda$ to keep $\epsilon_{N_i}$ fixed. As shown in the lower panel of Fig. \ref{fig:common-ps} for $T_{\rm RH}=0.1~ v_\phi$ and $f=0.005$, $\Lambda\gtrsim 6\times 10^{16}$ GeV is required along the black patches to satisfy the correct resonance condition. Consequently, for these points, higher values of $m_\chi$ ($\mathcal{O}(0.01~{\rm GeV})-\mathcal{O}({\rm GeV})$) as well as $T_{\rm RH}$ (that translates to a larger values of RHN masses\footnote{Decay of such heavy RHNs may emit gravitational waves via {\it bremsstrahlung} during leptogenesis \cite{Datta:2024tne}.}, in the range $\mathcal{O}(10^9~{\rm GeV})-\mathcal{O}(10^{12}~{\rm GeV})$) are needed to satisfy correct relic density.
	\begin{figure}[t]
		\centering
		\includegraphics[scale=0.54]{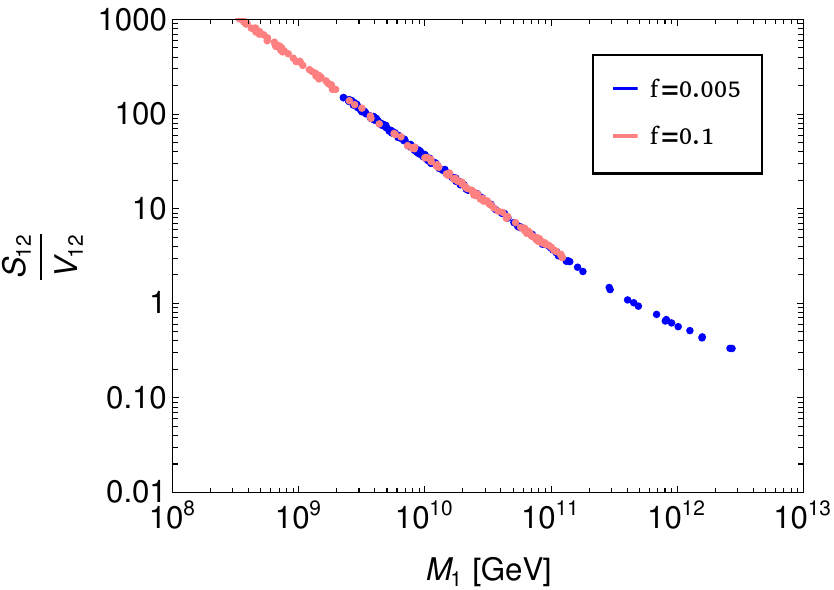}
		\caption{Plot showing the relative importance between the self energy and vertex contributions against $M_1$ for two choices of $f$. For $f =0.1$, it stops around $M_1 \sim 10^{11} $ GeV.}
		\label{SV-ratio}
	\end{figure}

	The requirement of satisfying the correct BAU boils down to generating a CP asymmetry of order $\epsilon_N \gtrsim 7 \times 10^{-6}$ 
	as the washout ($\eta$) of the produced lepton asymmetry via inverse decay is limited roughly by $\eta \lesssim 10^{-3}$, in terms of the DM parameter space\footnote{The actual contribution toward washout is however evaluated using the Boltzmann equations.}. This excludes the DM relic satisfied parameter space below the black patch with {$\theta_R = 0.78 + 0.42i$}. 
	We observe that for {$\theta_R = 0.78 + 0.42i$},
	the BAU satisfied black patch corresponds to a specific $\Lambda$ value for the upper panel with $f=0.1$ and $T_{\rm RH}/v_{\phi} = 0.5$ while for the lower panel, such a patch exhibits similar pattern till a point (close to $T_{\rm RH} \sim 10^{12}$ GeV or $M_1 \sim 10^{11}$ GeV) beyond which it shows a $\Lambda$ dependence. The same pattern is observed for different choices of $\theta_R$ as indicated in the Fig. \ref{fig:common-ps}. {However, choosing a $\theta_R$ other than the benchmark value $0.78 + 0.42i$ (which maximises the CP asymmetry) would restrict the parameter space from leptogenesis point of view as the amount of enhancement due to resonant leptogenesis is limited with those values of $\theta_R$.} This is actually reminiscent of the two distinct regimes of $\epsilon_{N}$ where the self energy domination over the vertex contribution characterises the fixed-$\Lambda$ regime (the range remains insensitive to $v_{\phi}$) while the $\Lambda$-dependent range stands for vertex correction dominated era. To make it more explicit, we include the Fig. \ref{SV-ratio}, where the relative strength of the self energy to vertex contributions in $\epsilon_{N_1}$ is depicted against the variation of RHN mass $M_1$. For $f=0.1$, it stops around $M_{1,2} \sim \mathcal{O}(10^{11})$ GeV signifying that $M_1$ is limited by such a value from the DM satisfied parameter space for $\Lambda\sim1.5 \times 10^{14}$ GeV. However, for the smaller choice of $f$, the DM parameter space allows for a higher $M_1$, thereby resulting into the probe of $\Lambda$-dependent regime in this case. So, overall, it is seen that the 1-loop vertex correction starts contributing significantly for $M_1\sim M_2 ~{\gtrsim}~ \mathcal{O}(10^{11})$ GeV. Therefore, we find that the study establishes a one-to-one correspondence between masses of the Majoron as DM and the RHN mass responsible for neutrino mass generation and BAU, as indicated in Fig. \ref{fig:MN-mchi}, thereby providing an intricate link between the two apparently disconnected problems of particle physics and cosmology. {Note that the typical behaviour of the bottom plot, above and below $M_1 \sim 10^{11}$ GeV, is suggestive of the characteristic switchover from the self energy domination to the vertex correction dominated phase.}

	\section{Summary and Conclusion}
	\label{conclusions}
	
	In this work, we have studied a Majoron model, extending the SM gauge symmetry with a global lepton number symmetry $U(1)_L$, augmented by a discrete $Z_2$ symmetry, including two RHNs and a scalar responsible for SSB of $U(1)_L$. 
	While the spontaneous breaking of the $U(1)_L$ naturally generates two exactly degenerate right handed neutrinos via the tree level terms respecting the lepton number symmetry, a tiny explicit lepton number breaking (but $Z_2$ symmetric) terms of dimension-5 breaks the degeneracy. The RHNs are not only responsible for generating the light neutrino masses via type I seesaw, but also contribute dominantly to the Majoron production in the early Universe. The quasi-degenerate RHNs allow an explanation of the baryon asymmetry of the Universe via resonant leptogenesis. It also opens up an interesting correlation between the Majoron as DM and BAU, not hitherto explored in the literature. 
	
	{The construction of the present Majoron model however includes a few assumptions such as imposing discrete symmetries and typical hierarchies among mass scales, which pave the way in obtaining such correlation. To begin with, we consider the explicit $U(1)_L$ breaking to take place only by higher dimensional operators. The presence of an underlying $Z_2$ symmetry not only restricts number of such operators to appear, but also helps in obtaining exactly degenerate pair of RHNs in absence of $U(1)_L$ explicit breaking terms. In continuation, a uniform coupling (or same cut-off scale) of the $U(1)_L$ breaking terms quadratic in RHNs, contributing toward the mass splitting $\Delta M$, is considered without loss of generality. 
		On the other hand, a CP symmetry under which $\Phi \rightarrow \Phi^*$ is present which ensures the dimension-5 Yukawa interaction of the second RHN to couple with $\Phi$ and $\Phi^*$ uniformly. We further consider the ordering of mass scales, $v_\phi>T_{\rm RH}>M_i$, in order to achieve correct DM phenomenology as well as in realizing leptogenesis. }
	
	
	
	We analysed the parameter space satisfying both the DM relic density and the BAU. The reheating temperature and the (effective) cut-off scale, signifying the explicit breaking of the global $U(1)_L$ symmetry, are bounded by $10^{15}$ GeV and Planck scale respectively. The analysis suggests a sub-GeV Majoron relevant to monochromatic neutrino search experiments, together RHNs in the mass range from intermediate ($10^6$ GeV) to high ($10^{13}$ GeV) scales. Unlike typical resonant leptogenesis occurring at the TeV scale, the high-scale resonant leptogenesis here has a characteristic dependence of self energy and vertex contributions to the CP asymmetry for RHN masses around $\mathcal{O}(10^{11})$ GeV.
	
	Overall, the model considered here not only provides a mechanism for generating tiny neutrino masses but also relates the production of dark matter with the baryon asymmetry of the Universe, offering a unified approach to resolve some of the most fundamental questions in particle physics and cosmology.

	\section*{Acknowledgements}
	RR and SFK acknowledge financial support from the STFC Consolidated Grant ST/X000583/1.
	SFK acknowledges the European Union's Horizon 2020 Research and Innovation programme under Marie Sklodowska-Curie grant agreement HIDDeN European ITN project (H2020-MSCA-ITN-2019//860881-HIDDeN). The work of AS is supported by the grants CRG/2021/005080 and MTR/2021/000774 from SERB, Govt. of India. AS also acknowledges the hospitality of the University of Southampton during a visit where the problem was formulated. SKM acknowledges the financial support received from the grant CRG/2021/005080 from SERB, Govt. of India.

	\appendix
	{ \section{CP violating couplings}
		\label{apndx_A}
		In this appendix, we discuss what happens if the CP symmetry is not conserved, in other words, unequal couplings of $\Phi$ and $\Phi^*$ in the Yukawa interaction of Eq. 6 are present. For example, one can make the Yukawa interaction of Eq. 6 non-uniform between $\Phi$ and $\Phi^*$ by considering an additional parameter $\alpha$ as 
		\beeq
		y_{\alpha 2}\overline{L_\alpha} \tilde{H}\mathcal{N}_2 \frac{(\Phi+\alpha \Phi^*)}{\Lambda}\supset i y_{\alpha 2}\overline{L_\alpha} \tilde{H}\mathcal{N}_2  \frac{(1-\alpha)}{\sqrt{2}\Lambda} \chi, 
		\eeq
		which induces additional Majoron production channel such as $LH\to N\chi$. However, the presence of the Yukawa coupling $y_{\alpha 2} \lesssim \mathcal{O}(1)$ in front along with $(1-\alpha) < 1$ suggests that this contribution would remain comparable or less compared to main production channel of Majoron $N_i N_i \to \chi\chi$. 
		
		On the other hand, the presence of $\alpha$ also enables Majoron to decay (hence, may cause instability) into two light neutrinos via active-sterile mixing. The corresponding decay width can be approximated as
		\beeq
		\Gamma_{\chi\to\nu \nu}\approx \left[ \frac{(1-\alpha)~ m_\nu}{\Lambda}\right]^2 \frac{m_\chi}{8\pi},
		\eeq
		which turns out to be suppressed compared to the similar decay of Majoron (already used to constrain the parameter space) as in Eq. 20, due to the additional suppression resulting from the appearance of $(1-\alpha)^2$ in numerator and 
		$\Lambda~ (> v_{\phi})$ in the denominator. Hence, in nutshell, the CP symmetry considered for $\Phi$ seems useful in restricting Majorons not to communicate with the $N_2$ (via Yukawa interaction of dimension-5) and the SM lepton sector (first two terms of Eq. 6 as well). }
	
	\bibliographystyle{apsrev4-1}
	\bibliography{ref}
\end{document}